\documentclass[12pt,thmsa]{article}
\usepackage{sw20jart}

\input{tcilatex}
\begin{document}

\author{Philip\ Maymin\thanks{%
The author's email address is \texttt{pzmaymin@fas.harvard.edu}.} \\
%EndAName
Harvard University}
\date{February 25, 1997}
\title{The lambda-q calculus can efficiently simulate quantum computers}
\maketitle

\begin{abstract}
We show that the lambda-q calculus can efficiently simulate quantum Turing
machines\ by showing how the lambda-q calculus can efficiently simulate a
class of quantum cellular automaton that are equivalent to quantum Turing
machines. We conclude by noting that the lambda-q calculus may be strictly
stronger than quantum computers because NP-complete problems such as
satisfiability are efficiently solvable in the lambda-q calculus but there
is a widespread doubt that they are efficiently solvable by quantum
computers.
\end{abstract}

\section{Introduction}

We show that the $\lambda ^{q}$-calculus defined in \cite{Maymin 1996} can
efficiently simulate the \emph{one-dimensional partitioned quantum cellular
automata} (1d-PQCA) defined in \cite{Watrous 1995}. By the equivalence of
1d-PQCA and quantum Turing machines\ (QTM) proved in \cite{Watrous 1995},
the $\lambda ^{q}$-calculus can efficiently simulate QTM.

We assume familiarity with both the $\lambda ^{q}$-calculus and 1d-PQCA as
defined in the above papers.

\section{Simulation}

To show that 1d-PQCA can be efficiently simulated by the $\lambda ^{q}$%
-calculus, we need to exhibit a $\lambda ^{q}$-term $M$ for a given 1d-PQCA $%
A$ such that $A$ after $k$ steps is in the same superposition as $M$ after $%
P\left( k\right) $ steps, with $P$ a polynomial.

We assume for now that the 1d-PQCA has transition amplitudes not over the
complex numbers, but over the positive and negative rationals. It has been
shown \cite{bernstein/vazirani} that this is equivalent to the general model
in QTM.

To express $A$ in $M$, we need to do the following things.

\begin{enumerate}
\item  Translate states of $A$ into $\lambda ^{q}$-terms that can be
compared\ (e.g. into Church numerals).

\item  Translate the acceptance states and the integer denoting the
acceptance cell into $\lambda ^{q}$-terms.

\item  Create a $\lambda ^{q}$-term $\mathbf{P}$ to mimic the operation of
the permutation $\sigma .$

\item  Translate the local transition function into a transition term. For
1d-PQCA this means translating the matrix $\Lambda $ into a term $\mathbf{L}$
comparing the initial state with each of the possible states and returning
the appropriate superposition.

\item  Determine an injective mapping of configurations of $A$ and
configurations of $M$.
\end{enumerate}

Although we will not write down $M$ in full, we note that within $M$ are the
mechanisms described above that take a single configuration, apply $\mathbf{P%
}$, and return the superposition as described by $\mathbf{L}.$

We recall that the contextual closure of the $\beta ^{q}$-relation is such
that $M,N\rightarrow _{\beta }M^{\prime },N^{\prime }$ where $M\rightarrow
_{\beta }M^{\prime }$and $N\rightarrow _{\beta }N^{\prime }.$ Thus there is
parallel reduction within superpositions. By inspection of the mechanisms
above it follows that $k$ steps of $A$ is equivalent to a polynomial of $k$
steps of $M$.

Steps 1, 2, and 3 are easy. We will use the following abbreviatory notation
for $\lambda ^{q}$-superpositions. We let $\left[ \left( M_{i}:n_{i}\right)
\right] $ be a rewriting of the term $\left[ N_{i}^{i\in I}\right] $ such
each of the $M_{i}$ are distinct and the integer $n_{i}$ represents the
count of each $M_{i}.$ We can also write this as $\left[ \left(
M_{i}:a_{i},b_{i},n_{i}\right) \right] $ such that $M_{i}\not{\equiv}M_{j}$
and $M_{i}\not{\equiv}\overline{M_{j}}$ for $i\neq j,$ all of the $M_{i}$
are of positive sign, the integer $a_{i}$ denotes the count of $M_{i},$ the
integer $b_{i}$ denotes the count of $\overline{M_{i}},$ and $%
n_{i}=a_{i}-b_{i}.$

Then for step 5, the $\lambda ^{q}$-superposition $\left[ \left(
M_{i}:a_{i},b_{i},n_{i}\right) \right] $ (let $n=\sum n_{i}$) will be
equivalent to the 1d-PQCA-superposition $\sum \frac{n_{i}}{n}\left| c\left(
M_{i}\right) \right\rangle ,$ where $c$ takes $\lambda ^{q}$-terms and
translates them into 1d-PQCA configurations. Essentially this means
stripping off everything other than the data, that is to say, the structure
containing the contents. Note that $c$ is not itself a $\lambda ^{q}$-term.
It merely performs a fixed syntactic operation, removing extraneous
information such as $\mathbf{P}$ and $\mathbf{L,}$ and translating the
Church numerals that represent states into the 1d-PQCA states. This is
injective because the mapping from states of $A$ into numerals is injective.
Thus, step 5 is complete.

Step 4 requires translating the $\Lambda $ matrix into a matrix of whole
numbers, and translating an arbitrary 1d--PQCA superposition into a $\lambda
^{q}$-superposition. The latter is done merely by multiplying each of the
amplitudes by the product of the denominators of all of the amplitudes, to
get integers. We call the product of the denominators here $d$. We perform a
similar act on the $\Lambda $ matrix, multiplying each element by the
product of all of the denominators of $\Lambda .$ We call this constant $b.$
Then we have that $T=b\Lambda $ is a matrix over integers. This matrix can
be considered notation for the $\lambda ^{q}$-term that checks if a given
state is a particular state and returns the appropriate superposition. For
instance, if 
\[
\Lambda =\left( 
\begin{array}{ll}
\frac{2}{3} & \frac{1}{3} \\ 
0 & 1
\end{array}
\right) 
\]
then 
\[
T=b\Lambda =9\Lambda =\left( 
\begin{array}{ll}
6 & 3 \\ 
0 & 9
\end{array}
\right) 
\]
which we can consider as alternate notation for 
\begin{eqnarray*}
\mathbf{Q} &\equiv &\lambda s.\text{\textbf{\ IF }(\textbf{EQUAL }}s\text{%
\textbf{1}) (\textbf{1,1,1,1,1,1,2,2,2)}} \\
&&\text{(\textbf{IF} (\textbf{EQUAL\ }}s\text{\textbf{2}) (\textbf{%
2,2,2,2,2,2,2,2,2}))}
\end{eqnarray*}
Then it follows that if $c$ is a superposition of configuration of $A$,
applying $\Lambda $ $k$ times results in the same superposition as applying $%
T$ $k$ times to the representation of $c$ in the $\lambda ^{q}$-calculus.

\section{Conclusion}

The $\lambda ^{q}$-calculus can efficiently simulate 1d-PQCA, which can
efficiently simulate QTM. Therefore the $\lambda ^{q}$-calculus can
efficiently simulate QTM. However, the $\lambda ^{q}$-calculus can
efficiently solve NP-complete problems such as satisfiability \cite{Maymin
1996}, while there is widespread belief (e.g. \cite{bennett}) that QTM\
cannot efficiently solve satisfiability. Thus, the greater the doubt that
QTM cannot solve NP-complete problems, the greater the justification in
believing that the $\lambda ^{q}$-calculus is strictly stronger than QTM.

\end{document}